\documentclass[dvips,preprint]{imsart}

\RequirePackage[OT1]{fontenc}
\usepackage{amsthm,amsmath}
\usepackage[numbers]{natbib}
\RequirePackage{hypernat}

\usepackage{amstext,amssymb,graphicx,rotating,array}
\usepackage{booktabs,colortbl,url}
\usepackage{algorithmic,algorithm}

\newcommand{\bflambda}{\boldsymbol{\lambda}}
\newcommand{\bftheta}{\boldsymbol{\theta}}
\newcommand{\bfX}{{\bf X}}
\newcommand{\bfV}{{\bf V}}
\newcommand{\bfW}{{\bf W}}

\startlocaldefs
\numberwithin{equation}{section}
\theoremstyle{plain}

\endlocaldefs

\begin{document}
\sloppy

\begin{frontmatter}
\title{Graphics Processing Units and High-Dimensional Optimization}
\runtitle{GPUs and High-Dimensional Optimization}

\begin{aug}
\author{\fnms{Hua} \snm{Zhou}\ead[label=e1]{huazhou@ucla.edu}},
\author{\fnms{Kenneth} \snm{Lange}\ead[label=e2]{klange@ucla.edu}}
\and
\author{\fnms{Marc A.} \snm{Suchard}\ead[label=e3]{msuchard@ucla.edu}}

\runauthor{Zhou, Lange, and Suchard}

\affiliation{University of California, Los Angeles}

\address{Department of Human Genetics, University of California, Los Angeles, \printead{e1}.}
\address{Departments of Biomathematics, Human Genetics, and Statistics, University of California,, Los Angeles, \printead{e2}.}
\address{Departments of Biomathematics, Biostatistics, and Human Genetics, University of California, Los Angeles, \printead{e3}.}
\end{aug}

\begin{abstract}
This paper discusses the potential of graphics processing units (GPUs) in high-dimensional
optimization problems. A single GPU card with hundreds of arithmetic cores can be inserted 
in a personal computer and dramatically accelerates many statistical algorithms. To exploit 
these devices fully, optimization algorithms should reduce to multiple parallel tasks, each accessing  
a limited amount of data.  These criteria favor EM and MM algorithms that separate parameters 
and data.  To a lesser extent block relaxation and coordinate descent and ascent also
qualify. We demonstrate the utility of GPUs in nonnegative matrix factorization, PET image reconstruction, and multidimensional scaling. Speedups
of 100 fold can easily be attained. Over the next decade, GPUs will fundamentally alter
the landscape of computational statistics. It is time for more statisticians to get
on-board.
\end{abstract}


\end{frontmatter}

\section{Introduction}

Statisticians, like all scientists, are acutely aware that the clock speeds on their desktops and laptops have stalled. Does this mean that statistical computing has hit a wall? The answer fortunately is no, but the hardware advances that we routinely expect have taken an interesting detour. Most computers now sold have two to eight processing cores. Think of these as separate CPUs on the same chip. Naive programmers rely on sequential algorithms and often fail to take advantage of more than a single core.  Sophisticated programmers, the kind who work for commercial firms such as Matlab, eagerly exploit parallel programming. However, multicore CPUs do not represent the only road to the success of statistical computing.

Graphics processing units (GPUs) have caught the scientific community by surprise. These devices are designed for graphics rendering in computer animation and games. Propelled by these nonscientific markets, the old technology of numerical (array) coprocessors has advanced rapidly. Highly parallel GPUs are now making computational inroads against traditional CPUs in image processing, protein folding, stock options pricing, robotics, oil exploration, data mining, and many other areas \cite{Owens:2007:ASO}.  We are starting to see orders of magnitude improvement on some hard computational problems.  Three companies, Intel, NVIDIA, and AMD/ATI, dominate the market. Intel is struggling to keep up with its more nimble competitors.

Modern GPUs support more vector and matrix operations, stream data faster, and possess more local memory per core than their predecessors.  They are also readily available as commodity items that can be inserted as video cards on modern PCs. GPUs have been criticized for their hostile programming environment and lack of double precision arithmetic and error correction, but these faults are being rectified.  The CUDA programming environment \cite{cuda-book} for NVIDIA chips is now easing some of the programming chores. We could say more about near-term improvements, but most pronouncements would be obsolete within months.

Oddly, statisticians have been slow to embrace the new technology. \citet{silberstein08} first demonstrated the potential for GPUs in fitting simple Bayesian networks.  Recently \citet{suchard09} have seen greater than $100$-fold speed-ups in MCMC simulations in molecular phylogeny.  \citet{holmes-tech-report} and \citet{tibbets09} are following suit with Bayesian model fitting via particle filtering and slice sampling. Finally, work is under-way to port common data mining techniques such as hierarchical clustering and multi-factor dimensionality reduction onto GPUs \citep{sinnott-armstrong09}. These efforts constitute the first wave of an eventual flood of statistical and data mining applications. The porting of GPU tools into the R environment will undoubtedly accelerate the trend 
\cite{buckner09}.

Not all problems in computational statistics can benefit from GPUs. Sequential algorithms are resistant unless they can be broken into parallel pieces. Even parallel algorithms can be problematic if the entire range of data must be accessed by each GPU.  Because they have limited memory, GPUs are designed to operate on short streams of data. The greatest speedups occur when all of the GPUs on a card perform the same arithmetic operation simultaneously.  Effective applications of GPUs in optimization involves both separation of data and separation of parameters. 

In the current paper, we illustrate how GPUs can work hand in glove with the MM algorithm, a generalization of the EM algorithm. In many optimization problems, the MM algorithm explicitly separates parameters by replacing the objective function by a sum of surrogate functions, each of which involves a single parameter. Optimization of the one-dimensional surrogates can be accomplished by assigning each subproblem to a different core. Provided the different cores each access just a slice of the data, the parallel subproblems execute quickly.  By construction the new point in parameter space improves the value of the objective function. In other words, MM algorithms are iterative ascent or descent algorithms. If they are well designed, then they separate parameters in high-dimensional problems. This is where GPUs enter. They offer most of the benefits of distributed computer clusters at a fraction of the cost. For this reason alone, computational statisticians need to pay attention to GPUs.

\begin{figure}[htpb]
\begin{center}
$$
\begin{array}{cc}
\includegraphics[width=2.3in]{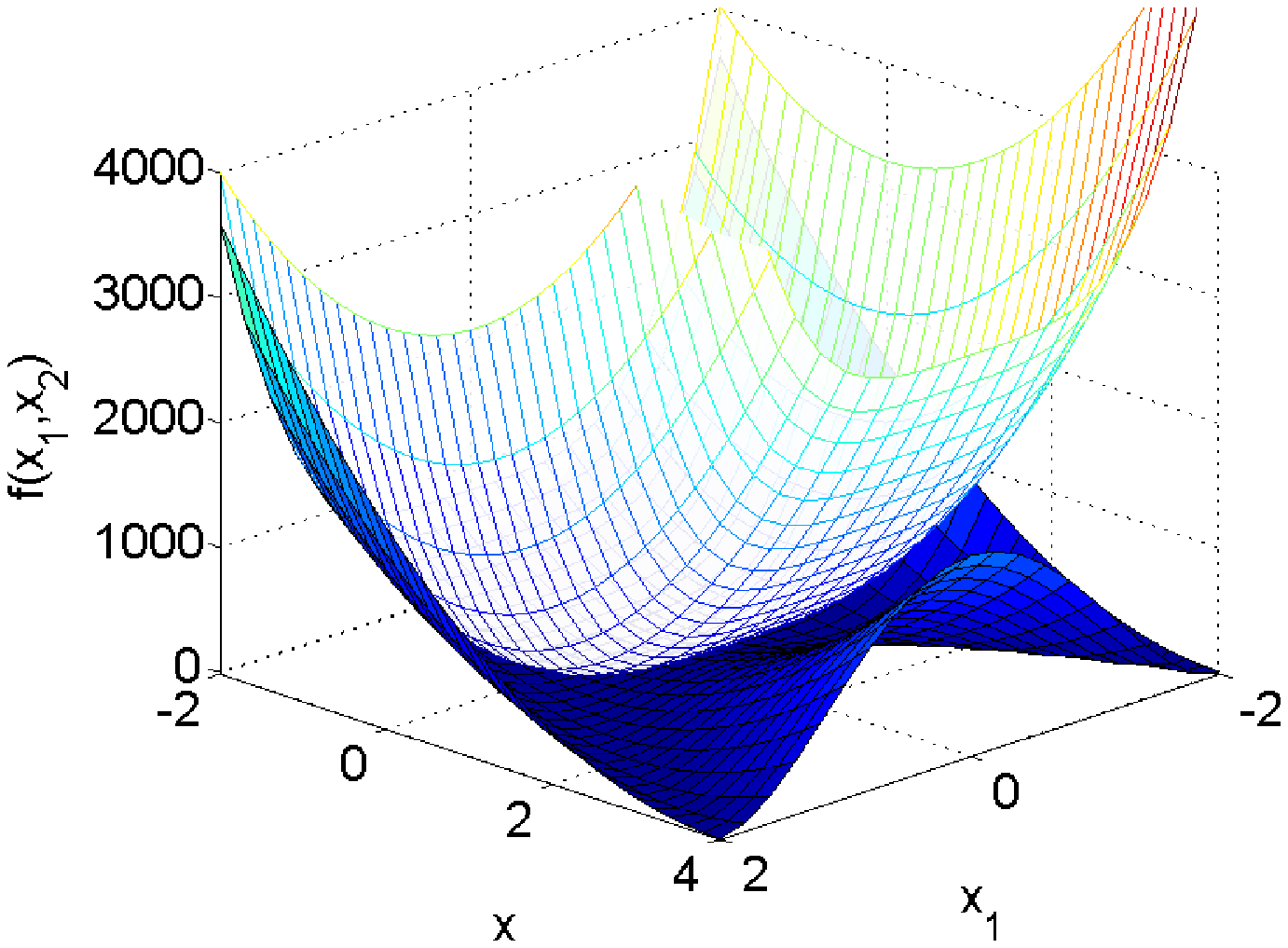} & \includegraphics[width=2.3in]{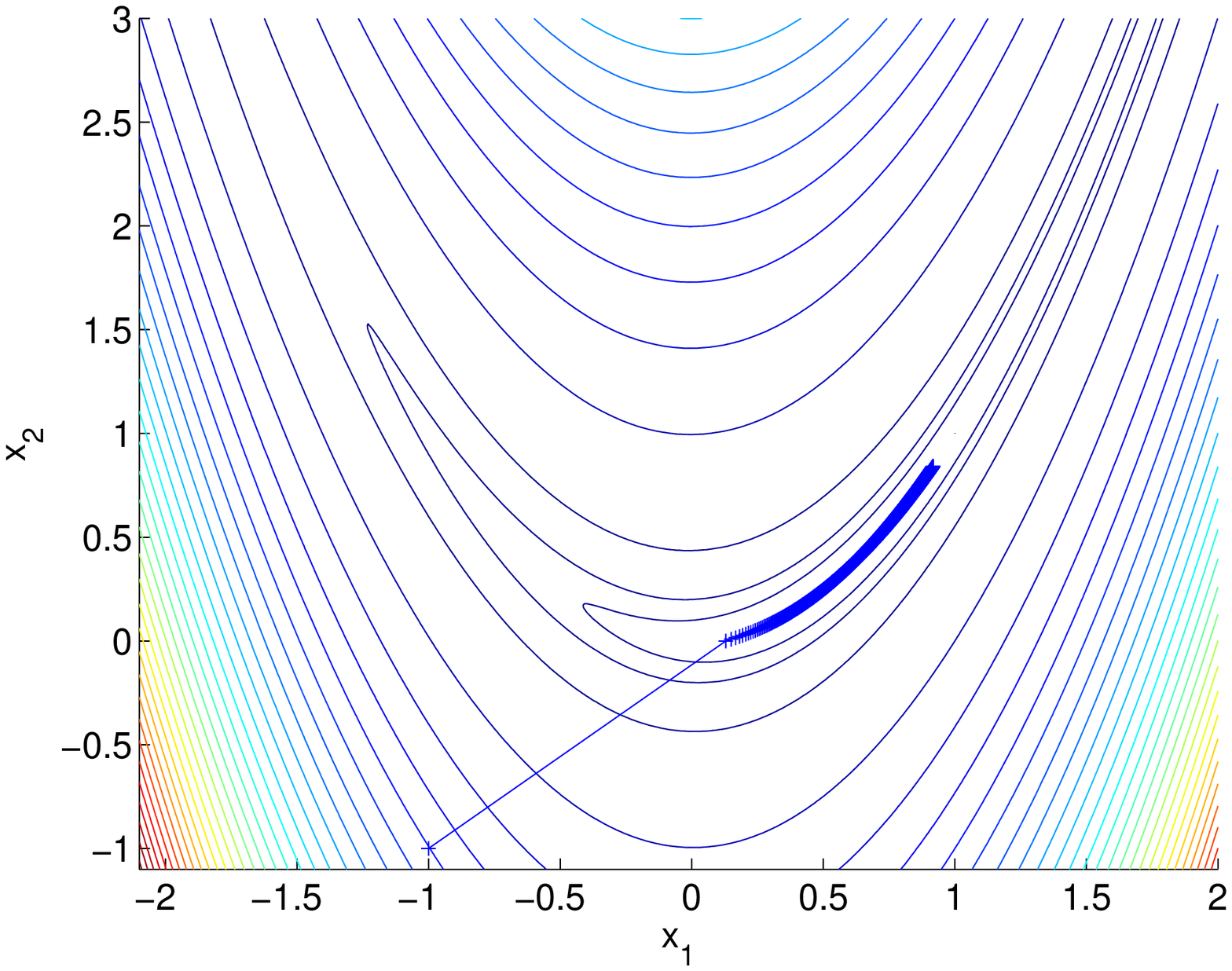}
\end{array}
$$
\end{center}
\caption{Left: The Rosenbrock (banana) function (the lower surface) and a majorization function at point (-1,-1) (the upper surface). Right: MM iterates.} 
\label{fig:Rosenbrock}
\end{figure}

Before formally defining the MM algorithm, it may help the reader to walk through a simple numerical example stripped of statistical content. Consider the Rosenbrock test function
\begin{eqnarray}
f({\bf  x}) & = & 100(x_1^2-x_2)^2+(x_1-1)^2 	\label{eqn:Rosenbrock-objective} \\
& = & 100(x_1^4 + x_2^2 - 2x_1^2 x_2) + (x_1^2 - 2x_1 + 1), \nonumber
\end{eqnarray}
familiar from the minimization literature. As we iterate toward the minimum at ${\bf x} ={\bf 1} = (1,1)$, we  construct a surrogate function that separates parameters. This is done by exploiting the obvious majorization
\begin{eqnarray*}
- 2 x_1^2 x_2 & \le & x_1^4 + x_2^2 + (x_{n1}^2 + x_{n2})^2 - 2 (x_{n1}^2 + x_{n2}) (x_1^2 + x_2),
\end{eqnarray*}
where equality holds when ${\bf x}$ and the current iterate ${\bf x}_n$ coincide. It follows that 
$f({\bf x})$ itself is majorized by the sum of the two surrogates
\begin{eqnarray*}
g_1(x_1 \mid {\bf x}_n) & = & 200 x_1^4 - [200( x_{n1}^2 + x_{n2})-1] x_1^2 - 2x_1 +1 \\
g_2(x_2 \mid {\bf x}_n) & = & 200 x_2^2 - 200( x_{n1}^2+x_{n2} ) x_2 + (x_{n1}^2 + x_{n2})^2.
\end{eqnarray*}
The left panel of Figure~\ref{fig:Rosenbrock} depicts the Rosenbrock function and its majorization
$g_1(x_1 \mid {\bf x}_n) + g_2(x_2 \mid {\bf x}_n)$ at the point $-{\bf 1}$. 

According to the MM recipe, at each iteration one must minimize the quartic polynomial $g_1(x_1 \mid {\bf x}_n)$ and the quadratic polynomial $g_2(x_2 \mid {\bf x}_n)$.  The quartic possesses either a single global minimum or two local minima separated by a local maximum These minima are the roots of the cubic function $g_1'(x_1|{\bf x}_n)$ and can be explicitly computed. We update $x_1$ by the root corresponding to the global minimum and $x_2$ via $x_{n+1,2} = \frac{1}{2}(x_{n1}^2+x_{n2})$. The right panel of Figure~\ref{fig:Rosenbrock} displays the iterates starting from ${\bf x}_0 = -{\bf 1}$. These immediately jump into the Rosenbrock valley and then slowly descend to ${\bf 1}$.  

Separation of parameters in this example makes it easy to decrease the objective function. This almost trivial advantage is amplified when we optimize functions depending on tens of thousands to millions of parameters. In these settings, Newton's method and variants such as Fisher's scoring are fatally handicapped by the need to store, compute, and invert huge Hessian or information matrices. On the negative side of the balance sheet, MM algorithms are often slow to converge. This disadvantage is usually outweighed by the speed of their updates even in sequential mode. If one can harness the power of parallel processing GPUs, then MM algorithms become the method of choice for many high-dimensional problems.

We conclude this introduction by sketching a roadmap to the rest of the paper. Section~\ref{sec:MM} reviews the MM algorithm.  Section~\ref{sec:examples} discusses three high-dimensional MM examples.  Although the algorithm in each case is known, we present brief derivations to illustrate how simple inequalities drive separation of parameters. We then implement each algorithm on a realistic problem and compare running times in sequential and parallel modes. We purposefully omit programming syntax since many tutorials already exist for this purpose, and material of this sort is bound to be ephemeral. Section \ref{discussion_section} concludes with a brief discussion of other statistical applications of GPUs and other methods of accelerating optimization algorithms.

\section{MM Algorithms}
\label{sec:MM}

The MM algorithm like the EM algorithm is a principle for creating optimization algorithms. In minimization the acronym MM stands for majorization-minimization; in maximization it stands for minorization-maximization.  Both versions are convenient in statistics. For the moment we will concentrate on maximization.

Let $f(\bftheta)$ be the objective function whose maximum we seek. Its argument $\bftheta$ can be high-dimensional and vary over a constrained subset $\Theta$ of Euclidean space. An MM algorithm involves minorizing $f(\bftheta)$ by a surrogate function $g(\bftheta \mid \bftheta_n)$ anchored at the current iterate $\bftheta_n$ of the search. The subscript $n$ indicates iteration number throughout this article. If $\bftheta_{n+1}$ denotes the maximum of $g(\bftheta \mid \bftheta_n)$ with respect to its left argument, then the MM principle declares that $\bftheta_{n+1}$ increases 
$f(\bftheta)$ as well. Thus, MM algorithms revolve around a basic ascent property. 

Minorization is defined by the two properties
\begin{eqnarray}
f(\bftheta_n) & = & g( \bftheta_n \mid \bftheta_n)  \label{minorization_definition1} \\
f(\bftheta) & \ge & g(\bftheta \mid \bftheta_n)\: , \quad \quad \bftheta \ne \bftheta_n . 
\label{minorization_definition2}
\end{eqnarray}
In other words, the surface $\bftheta \mapsto g(\bftheta \mid \bftheta_n)$ lies below the surface $\bftheta \mapsto f(\bftheta)$ and is tangent to it at the point $\bftheta=\bftheta_n$. Construction of the minorizing function $g(\bftheta \mid \bftheta_n)$ constitutes the first M of the MM algorithm. In our examples 
$g(\bftheta \mid \bftheta_n)$ is chosen to separate parameters. 

In the second M of the MM algorithm, one maximizes the surrogate $g(\bftheta \mid \bftheta_n)$ rather than 
$f(\bftheta)$ directly. It is straightforward to show that the maximum point $\bftheta_{n+1}$ satisfies the ascent property $f(\bftheta_{n+1}) \ge f(\bftheta_n)$.  The proof 
\begin{eqnarray*}
f(\bftheta_{n+1}) & \ge & g(\bftheta_{n+1} \mid \bftheta_n) \;\; \ge \;\; g(\bftheta_n \mid \bftheta_n)
\;\; = \;\;  f(\bftheta_n)
\end{eqnarray*}
reflects definitions (\ref{minorization_definition1}) and (\ref{minorization_definition2}) and the choice of $\bftheta_{n+1}$.  The ascent property is the source of the MM algorithm's numerical stability
and remains valid if we merely increase $g(\bftheta \mid \bftheta_n)$ rather than maximize it. In many problems MM updates are delightfully simple to code, intuitively compelling, and automatically consistent with parameter constraints. In minimization we seek a majorizing function $g(\bftheta \mid \bftheta_n)$ lying above the surface $\bftheta \mapsto f(\bftheta)$ and tangent to it at the point $\bftheta=\bftheta_n$.  Minimizing $g(\bftheta \mid \bftheta_n)$ drives $f(\bftheta)$ downhill. 

The celebrated Expectation-Maximization (EM) algorithm \cite{Dempster77EM,McLachlan08EMBook} is a special case of the MM algorithm. The $Q$-function produced in the E step of the EM algorithm constitutes a minorizing function of the loglikelihood.  Thus, both EM and MM share the same advantages: simplicity, stability, graceful adaptation to constraints, and the tendency to avoid large matrix inversion.  The more general MM perspective frees algorithm derivation from the missing data straitjacket and invites wider applications. For example, our multi-dimensional scaling (MDS) and non-negative matrix factorization (NNFM) examples involve no likelihood functions. \citet{WuLange09EMMM} briefly summarize the history of the MM algorithm and its relationship to the EM algorithm.   

The convergence properties of MM algorithms are well-known \citep{Lange04Optm}. In particular,
five properties of the objective function $f(\bftheta)$ and the MM algorithm map $\bftheta \mapsto M(\bftheta)$ guarantee convergence to a stationary point of $f(\bftheta)$: (a) $f(\bftheta)$ is coercive on its open domain; (b) $f(\bftheta)$ has only isolated stationary points; (c) $M(\bftheta)$ is continuous; (d) $\bftheta^*$ is a fixed point of $M(\bftheta)$ if and only if $\bftheta^*$ is a stationary point of 
$f(\bftheta)$; and (e) $f[M(\bftheta^*)] \ge f(\bftheta^*)$, with equality if and only if $\bftheta^*$ is a fixed point of $M(\bftheta)$. These conditions are easy to verify in many applications. The local rate
of convergence of an MM algorithm is intimately tied to how well the surrogate function 
$g(\bftheta \mid \bftheta^*)$ approximates the objective function $f(\bftheta)$ near the optimal point 
$\bftheta^*$.

\section{Numerical Examples}
\label{sec:examples}

In this section, we compare the performances of the CPU and GPU implementations of three classical MM algorithms coded in C++: (a) non-negative matrix factorization (NNMF), (b) positron emission tomography (PET), and (c) multidimensional scaling (MDS). In each case we briefly derive the algorithm from the MM perspective. For the CPU version, we iterate until the relative change 
\begin{eqnarray*}
\frac{|f(\bftheta_n)-f(\bftheta_{n-1})|}{|f(\bftheta_{n-1})|+1}
\end{eqnarray*}
of the objective function $f(\bftheta)$ between successive iterations falls below a pre-set threshold $\epsilon$ or the number of iterations reaches a pre-set number $n_{\rm max}$, whichever comes first. In these examples, we take $\epsilon = 10^{-9}$ and $n_{\rm max}=100,000$. For ease of comparison, we iterate the GPU version for the same number of steps as the CPU version. Overall, we see anywhere from a 22-fold to 112-fold decrease in total run time. The source code is freely available from the first author.

Table~\ref{table:system-config} shows how our desktop system is configured. Although the CPU is a high-end processor with four cores, we use just one of these for ease of comparison. In practice, it takes considerable effort to load balance the various algorithms across multiple CPU cores. With 240 GPU cores, the GTX 280 GPU card delivers a peak performance of about 933 GFlops in single precision. This card is already obsolete. Newer cards possess twice as many cores, and up to four cards can fit inside a single desktop computer. It is relatively straightforward to program multiple GPUs. Because previous generation GPU hardware is largely limited to single precision, this is a worry in scientific computing. To assess the extent of roundoff error, we display the converged values of the objective functions to ten significant digits. Only rarely is the GPU value far off the CPU mark. Finally, the extra effort in programming the GPU version is relatively light. Exploiting the standard CUDA library \cite{cuda-book}, it takes 77, 176, and 163 extra lines of GPU code to implement the NNMF, PET, and MDS examples, respectively. 

\begin{table}
\begin{center}
\begin{tabular}{ccc}
\toprule
 & CPU & GPU	\\
\midrule
Model & Intel Core 2 &  NVIDIA GeForce \\
& Extreme X9440 & GTX 280	\\
\# Cores & 4 & 240	\\
Clock & 3.2G & 1.3G \\
Memory & 16G & 1G	\\
\bottomrule
\end{tabular}
\end{center}
\caption{Configuration of the desktop system}
\label{table:system-config}
\end{table}

\subsection{Non-Negative Matrix Factorizations}

Non-negative matrix factorization (NNMF) is an alternative to principle component analysis useful in modeling, compressing, and interpreting nonnegative data such as observational counts and images. The
articles \cite{LeeSeung99NNMF,LeeSeung01NNMFAlgo,Berry07NNMF} discuss in detail algorithm development and statistical applications of NNMF. The basic problem is to approximate a data matrix $\bfX$ with nonnegative entries $x_{ij}$ by a product $\bfV \bfW$ of two low rank matrices $\bfV$ and $\bfW$ with nonnegative entries $v_{ik}$ and $w_{kj}$. Here $\bfX$, $\bfV$, and $\bfW$ are $p \times q$, $p \times r$, and $r \times q$, respectively, with $r$ much smaller than $\min\{p,q\}$. One version of NNMF minimizes the objective function
\begin{eqnarray}
f(\bfV,\bfW) & = & \| \bfX - \bfV \bfW \|_{\text{F}}^2 \;\; = \;\;  
\sum_i \sum_j \Big( x_{ij} - \sum_k v_{ik} w_{kj} \Big)^2,  \label{eqn:nnmf-objfn}
\end{eqnarray}
where $\|\cdot\|_{\text{F}}$ denotes the Frobenius-norm.  To get an idea of the scale of NNFM imaging problems, $p$ (number of images) can range $10^1-10^4$, $q$ (number of pixels per image) can surpass $10^2-10^4$, and one seeks a rank $r$ approximation of about 50.  Notably, part of the winning solution of the Netflix challenge relies on variations of NNMF \cite{KorBel09}. For the Netflix data matrix, $p=480,000$ (raters), 
$q=18,000$ (movies), and $r$ ranged from 20 to 100.

Exploiting the convexity of the function $x \mapsto (x_{ij} - x)^2$, one can derive the inequality 
\begin{eqnarray*}
\Big( x_{ij} - \sum_k v_{ik} w_{kj} \Big)^2 & \le & \sum_k \frac{a_{nikj}}{b_{nij}} \left( x_{ij} - \frac{b_{nij}}{a_{nikj}} v_{ik} w_{kj} \right)^2
\end{eqnarray*}
where $a_{nikj} = v_{nik} w_{nkj}$ and $b_{nij} = \sum_k a_{nikj}$. This leads to the surrogate function
\begin{eqnarray}
g(\bfV,\bfW \mid \bfV_{n},\bfW_{n}) & = & \sum_i \sum_j \sum_k \frac{a_{nikj}}{b_{nij}} \left( x_{ij} - \frac{b_{nij}}{a_{nikj}} v_{ik} w_{kj} \right)^2
\label{frobenius_majorization}
\end{eqnarray}
majorizing the objective function $f(\bfV,\bfW) = \|\bfX-\bfV \bfW\|_{\text{F}}^2$. Although the majorization (\ref{frobenius_majorization}) does not achieve a complete separation of parameters, it does if we fix $\bfV$ and update $\bfW$ or vice versa. This strategy is called block relaxation.

If we elect to minimize $g(\bfV,\bfW \mid \bfV_{n},\bfW_{n})$ holding $\bfW$ fixed at $\bfW_n$, then the stationarity condition for $\bfV$ reads
\begin{eqnarray*}
\frac{\partial}{\partial v_{ik}} g(\bfV,\bfW_{n} \mid \bfV_{n},\bfW_{n}) & = &  -2\sum_j \Big( x_{ij} - \frac{b_{nij}}{a_{nikj}} v_{ik} w_{nkj} \Big) w_{nkj} \;\; = \;\; 0.
\end{eqnarray*}
Its solution furnishes the simple multiplicative update
\begin{eqnarray}
v_{n+1,ik} & = & v_{nik} \frac{\sum_j x_{ij} w_{nkj}}{\sum_j b_{nij} w_{nkj}}. \label{nnmf_v_update}
\end{eqnarray}
Likewise the stationary condition
\begin{eqnarray*}
\frac{\partial}{\partial w_{kj}} g(\bfV_{n+1},\bfW \mid \bfV_{n+1},\bfW_{n})& = & 0
\end{eqnarray*}
gives the multiplicative update
\begin{eqnarray}
w_{n+1,kj} & = & w_{nkj} \frac{\sum_i x_{ij} v_{n+1,ik}}{\sum_i c_{nij} v_{n+1,ik}}, \label{nnmf_w_update}
\end{eqnarray}
where $c_{nij} = \sum_k v_{n+1,ik}w_{nkj}$.  Close inspection of the multiplicative updates (\ref{nnmf_v_update}) and (\ref{nnmf_w_update}) shows that their numerators depend on the matrix products 
$\bfX \bfW_n^t$ and $\bfV_{n+1}^t \bfX $ and their denominators depend on the matrix products $\bfV_n \bfW_n \bfW_n^t$ and $\bfV_{n+1}^t \bfV_{n+1} \bfW_n$. Large matrix multiplications are very fast on GPUs because CUDA implements in parallel the BLAS (basic linear algebra subprograms) library widely applied in numerical analysis \cite{cublas-book08}. Once the relevant matrix products are available, each elementwise update of $v_{ik}$ or $w_{kj}$ involves just a single multiplication and division. These scalar operations are performed in parallel through hand-written GPU code. Algorithm~\ref{algo:nnmf-multiplicative} summarizes the steps in performing NNMF.

\begin{algorithm}
\begin{algorithmic}
\STATE Initialize: Draw $v_{0ik}$ and $w_{0kj}$ uniform on (0,1) for all $1 \le i \le p$, $1 \le k \le r$, $1 \le j \le q$
\REPEAT
\STATE Compute $\mathbf{X} \mathbf{W}_{n}^t$ and $\mathbf{V}_{n}\mathbf{W}_{n}\mathbf{W}_{n}^t$
\STATE $v_{n+1,ik} \leftarrow v_{nik} \cdot \{\mathbf{X} \mathbf{W}_{n}^t \}_{ik} \, / \, \{\mathbf{V}_{n}\mathbf{W}_{n}\mathbf{W}_{n}^t\}_{ik}$ for all $1 \le i \le p$, $1 \le k \le r$
\STATE Compute $\mathbf{V}_{n+1}^t \mathbf{X} $ and $\mathbf{V}_{n+1}^t\mathbf{V}_{n+1}\mathbf{W}_n$
\STATE $w_{n+1,kj} \leftarrow w_{nkj} \cdot \{\bfV_{n+1}^t \bfX\}_{kj} \, / \, \{\bfV_{n+1}^t\bfV_{n+1}\bfW_{n}\}_{kj}$ for all $1 \le k \le r$, $1 \le j \le q$
\UNTIL{convergence occurs}
\end{algorithmic}
\caption{(NNMF) Given $\bfX \in \mathbb{R}_+^{p \times q}$, find $\bfV \in \mathbb{R}_+^{p \times r}$ and $\bfW \in \mathbb{R}_+^{r \times q}$ minimizing $\|\bfX - \bfV \bfW\|_{\text{F}}^2$.}
\label{algo:nnmf-multiplicative}
\end{algorithm}

\begin{figure}
\begin{center}
\includegraphics[width=4.7in]{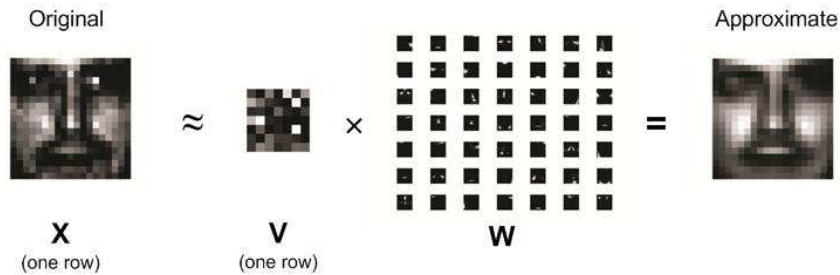}
\end{center}
\caption{Approximation of a face image by rank-49 NNMF: coefficients $\times$ basis images = approximate image.}
\label{fig:approx-face}
\end{figure}

We now compare CPU and GPU versions of the multiplicative NNMF algorithm on a training set of face images. Database \#1 from the MIT Center for Biological and Computational Learning (CBCL) \cite{MITCBCL} reduces to a matrix $\bfX$ containing $p=2,429$ gray scale face images with $q=19 \times 19 = 361$ pixels per face. Each image (row) is scaled to have mean and standard deviation 0.25. Figure
\ref{fig:approx-face} shows the recovery of the first face in the database using a rank $r=49$ decomposition. The 49 basis images (rows of $\bfW$) represent different aspects of a face. The rows of $\bfV$ contain the coefficients of these parts estimated for the various faces. Some of these facial features are immediately obvious in the reconstruction. Table~\ref{table:nnmf-faces} compares the run times of Algorithm~\ref{algo:nnmf-multiplicative} implemented on our CPU and GPU respectively.  We observe a 22 to 112-fold speed-up in the GPU implementation. Run times for the GPU version depend primarily on the number of iterations to convergence and very little on the rank $r$ of the approximation. Run times of the CPU version scale linearly in both the number of iterations and $r$.

\begin{table}
\begin{center}
\begin{tabular}{crrrrrc}
\toprule
& & \multicolumn{2}{c}{CPU} & \multicolumn{2}{c}{GPU} & \\ 
 \cmidrule(lr){3-4} \cmidrule(lr){5-6} 
Rank $r$ & Iters & Time & Function & Time & Function & Speedup \\ 
\midrule
10 & 25459    & 1203     & 106.2653503  &  55       & 106.2653504 & 22 \\
20 & 87801    & 7564     & 89.56601262  &  163      & 89.56601287 & 46 \\
30 & 55783    & 7013     & 78.42143486  &  103      & 78.42143507 & 68 \\
40 & 47775    & 7880     & 70.05415929  &  119      & 70.05415950 & 66 \\
50 & 53523    & 11108    & 63.51429261  &  121      & 63.51429219 & 92 \\
60 & 77321    & 19407    & 58.24854375  &  174      & 58.24854336 & 112 \\
\bottomrule
\end{tabular}
\end{center}
\caption{Run-time (in seconds) comparisons for NNMF on the MIT CBCL face image data. The dataset contains $p=2,429$ faces with $q = 19 \times 19 = 361$ pixels per face. The columns labeled Function refer to the converged value of the objective function.}
\label{table:nnmf-faces}
\end{table}

It is worth stressing a few points. First, the objective function~(\ref{eqn:nnmf-objfn}) is convex in $\bfV$ for $\bfW$ fixed, and vice versa but not jointly convex. Thus, even though the MM algorithm enjoys the descent property, it is not guaranteed to find the global minimum \citep{Berry07NNMF}.  There are two good alternatives to the multiplicative algorithm. First, pure block relaxation can be conducted by alternating least squares (ALS). In updating $\bfV$ with $\bfW$ fixed, ALS omits majorization and solves the $p$ separated nonnegative least square problems
\begin{eqnarray*}
	\min_{\bfV(i,:)} \|\bfX(i,:) - \bfV(i,:)\bfW]\|_2^2 \quad \text{ subject to } \bfV(i,:) \ge 0,
\end{eqnarray*}
where $\bfV(i,:)$ and $\bfX(i,:)$ denote the $i$-th row of the corresponding matrices. Similarly, in updating $\bfW$ with $\bfV$ fixed, ALS solves $q$ separated nonnegative least square problems. Another possibility is to change the objective function to
\begin{eqnarray*}
	L(\bfV,\bfW) & = & \sum_i \sum_j \Big[ x_{ij} \ln \Big( \sum_k v_{ik} w_{kj} \Big) - \sum_k v_{ik} w_{kj} \Big]
\end{eqnarray*}
according to a Poisson model for the counts $x_{ij}$ \cite{LeeSeung99NNMF}. This works even when some entries $x_{ij}$ fail to be integers, but the Poisson loglikelihood interpretation is lost. A pure MM algorithm for maximizing $L(\bfV,\bfW)$ is
\begin{eqnarray*}
v_{n+1,ik} & = & v_{nik} \sqrt{ \frac{\sum_j x_{ij} w_{nkj} / b_{nij}}{\sum_j w_{nkj}}  }, \quad w_{n+1,ij} = w_{nkj} \sqrt{ \frac{\sum_i x_{ij} v_{nik} / b_{nij}}{\sum_i v_{nik}}}.
\end{eqnarray*}
Derivation of these variants of Lee and Seung's \cite{LeeSeung99NNMF} Poisson updates is left to the reader.

\subsection{Positron Emission Tomography}

The field of computed tomography has exploited EM algorithms for many years. In positron emission tomography (PET), the reconstruction problem consists of estimating the Poisson emission intensities $\bflambda = (\lambda_1,\ldots,\lambda_p)$ of $p$ pixels arranged in a 2-dimensional grid surrounded by an array of photon detectors. The observed data are coincidence counts $(y_1, \ldots y_d)$ along $d$ lines of flight connecting pairs of photon detectors. The loglikelihood under the PET model is
\begin{eqnarray*}
L(\bflambda) & = & \sum_i \Big[y_i \ln\Big( \sum_j
e_{ij} \lambda_j\Big)-\sum_j e_{ij} \lambda_j\Big] ,
\end{eqnarray*}
where the $e_{ij}$ are constants derived from the geometry of the grid and the detectors. Without loss of generality, one can assume $\sum_i e_{ij}=1$ for each $j$.  It is straightforward to derive the traditional EM algorithm \cite{Lange84PET,Vardi85PET} from the MM perspective using the concavity of the function $\ln s$. Indeed, application of Jensen's inequality produces the minorization 
\begin{eqnarray*}
L(\bflambda) & \ge & \sum_i y_i \sum_j w_{nij}
\ln\Big( \frac{e_{ij} \lambda_j}{w_{nij}} \Big) -\sum_i \sum_j e_{ij} \lambda_j
\;\; = \;\; Q(\bflambda \mid \bflambda_n),
\end{eqnarray*}
where $w_{nij} = e_{ij} \lambda_{nj}/(\sum_k e_{ik} \lambda_{nk})$. This maneuver again separates parameters. The stationarity conditions for the surrogate $Q(\bflambda \mid \bflambda_n)$ supply the parallel updates
\begin{eqnarray}
\lambda_{n+1,j} & = & \frac{\sum_i y_i w_{nij}}{\sum_i e_{ij}}. \label{PET_algorithm}
\end{eqnarray}

The convergence of the PET algorithm (\ref{PET_algorithm}) is frustratingly slow, even under systematic acceleration \cite{Roland07PET,ZhouAlexanderLange09QN}. Furthermore, the reconstructed images are of poor quality with a grainy appearance. The early remedy of premature halting of the algorithm cuts computational cost but is entirely {\it ad hoc}, and the final image depends on initial conditions. A better option is add a roughness penalty to the loglikelihood. This device not only produces better images but also accelerates convergence.  Thus, we maximize the penalized loglikelihood
\begin{eqnarray}
f(\bflambda) & = & L(\bflambda) -\frac{\mu}{2} \sum_{\{j,k\}
\in {\cal N}} (\lambda_j - \lambda_k)^2 \label{eqn:PET-objpen}	
\end{eqnarray}
where $\mu$ is the roughness penalty constant, and ${\cal N}$ is the neighborhood system that pairs spatially adjacent pixels. An absolute value penalty is less likely to deter the formation of edges than a square penalty, but it is easier to deal with a square penalty analytically, and we adopt it for the sake of simplicity. In practice, visual inspection of the recovered images guides the selection of the roughness penalty constant $\mu$.

To maximize $f(\bflambda)$ by an MM algorithm, we must minorize the penalty in a manner consistent with the separation of parameters.  In view of the evenness and convexity of the function $s^2$, we have 
\begin{eqnarray*}
(\lambda_j - \lambda_k)^2 & \leq & \frac{1}{2} (2 \lambda_j -
\lambda_{nj} - \lambda_{nk})^2
+  \frac{1}{2} (2 \lambda_k - \lambda_{nj} - \lambda_{nk})^2 .
\end{eqnarray*}
Equality holds if $\lambda_j+\lambda_k = \lambda_{nj} + \lambda_{nk}$, which is true when $\bflambda  = \bflambda_n$. Combining our two minorizations furnishes the surrogate function
\begin{eqnarray*}
	g(\bflambda \mid \bflambda_n) = Q(\bflambda \mid \bflambda_n) -\frac{\mu}{4} 
\sum_{\{j,k\} \in {\cal N}}\Big[(2 \lambda_j - \lambda_{nj} - \lambda_{nk})^2 + (2 \lambda_k - \lambda_{nj} - \lambda_{nk})^2  \Big].
\end{eqnarray*}
To maximize $g(\bflambda \mid \bflambda_n)$, we define ${\cal N}_j = \{k: \{j,k\} \in {\cal N}\}$ and set the partial derivative
\begin{eqnarray}
\frac{\partial}{\partial \lambda_j} g(\bflambda \mid \bflambda_n) & = & \sum_i
\Big[\frac{y_i  w_{nij}}{\lambda_j}-e_{ij}\Big] - \mu \sum_{k: \in {\cal N}_j} (2
\lambda_j-\lambda_{nj} - \lambda_{nk})
   \label{eqn:next-lambda}
\end{eqnarray}
equal to 0 and solve for $\lambda_{n+1,j}$. Multiplying equation (\ref{eqn:next-lambda}) by $\lambda_j$ produces a quadratic with roots of opposite signs. We take the positive root
\begin{eqnarray*}
\lambda_{n+1,j} & = & \frac{-b_{nj} - \sqrt{b_{nj}^{2} - 4a_jc_{nj}}}{2a_j},
\end{eqnarray*}
where
\begin{eqnarray*}
a_{j} & = & -2 \mu \sum_{k \in {\cal N}_j} 1, \quad 
b_{nj} \;\; = \;\; \sum_{k \in {\cal N}_j} (\lambda_{nj}+\lambda_{nk}) -1, \quad
c_{nj} \;\; = \;\; \sum_i y_i w_{nij} .
\end{eqnarray*}
Algorithm \ref{algo:PET} summarizes the complete MM scheme. Obviously, complete parameter separation is crucial. The quantities $a_j$ can be computed once and stored. The quantities $b_{nj}$ and $c_{nj}$ are computed for each $j$ in parallel. To improve GPU performance in computing the sums over $i$, we exploit the widely available parallel sum-reduction techniques \citep{silberstein08}. Given these results, a specialized but simple GPU code computes the updates $\lambda_{n+1,j}$ for each $j$ in parallel.

\begin{figure}
\begin{center}
\includegraphics[width=2in]{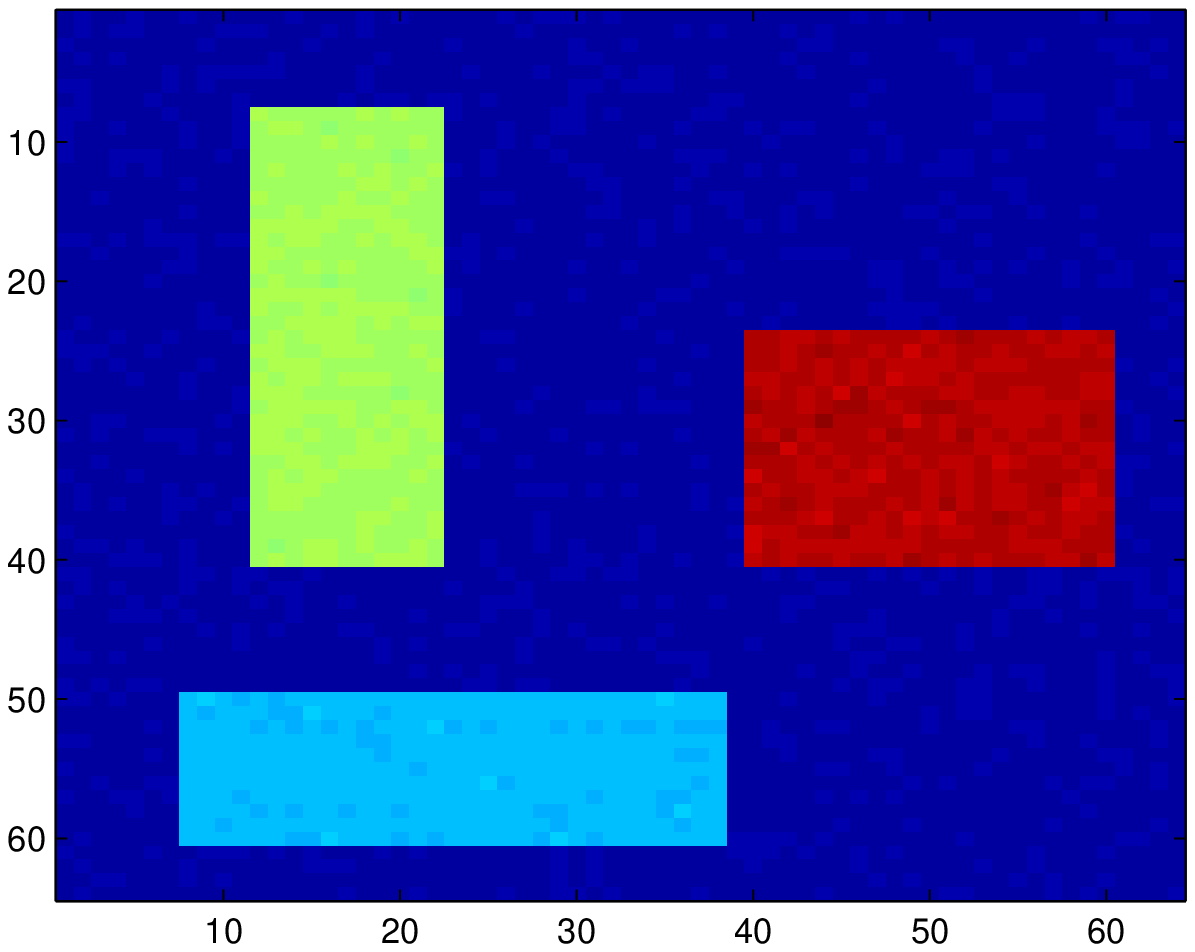}
$$
\begin{array}{cc}
	\includegraphics[width=2in]{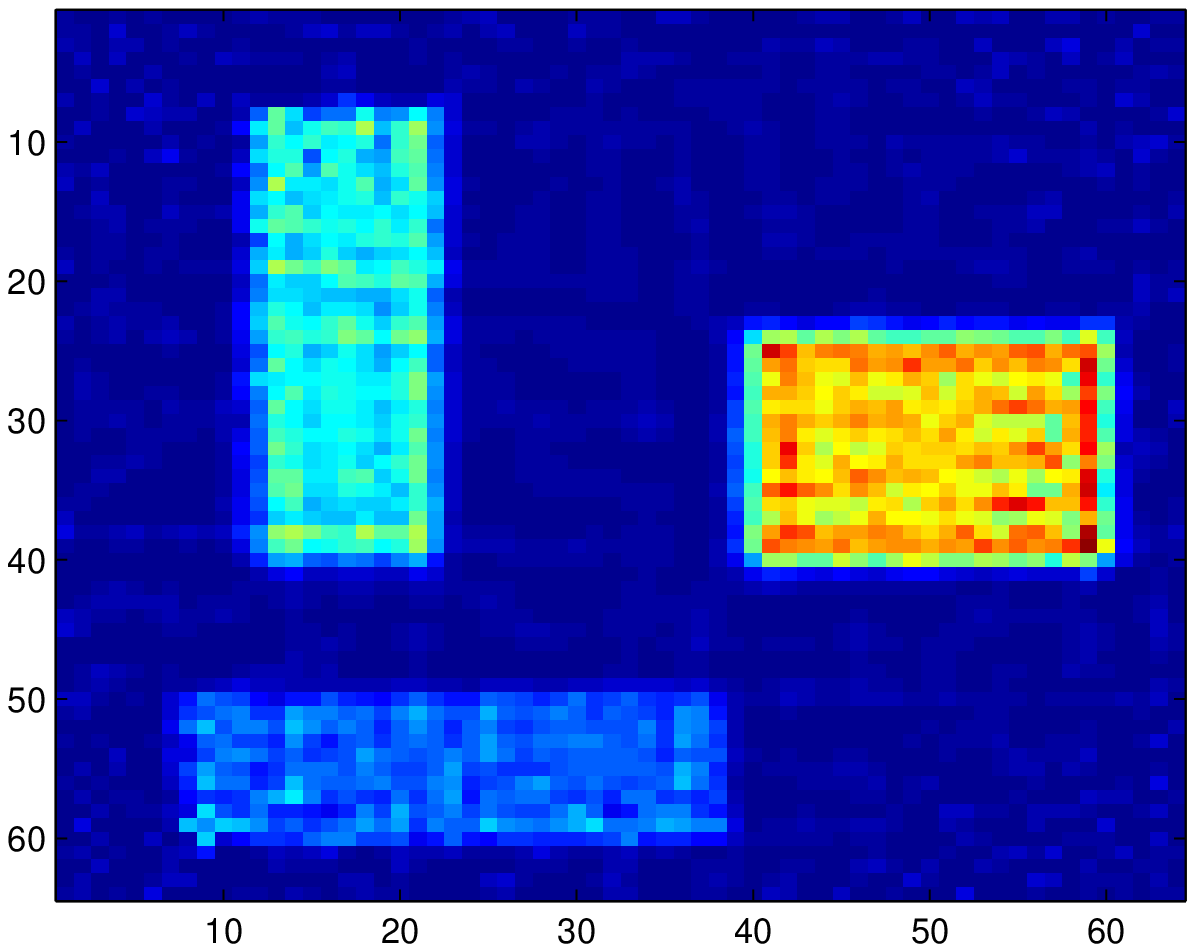} & \includegraphics[width=2in]{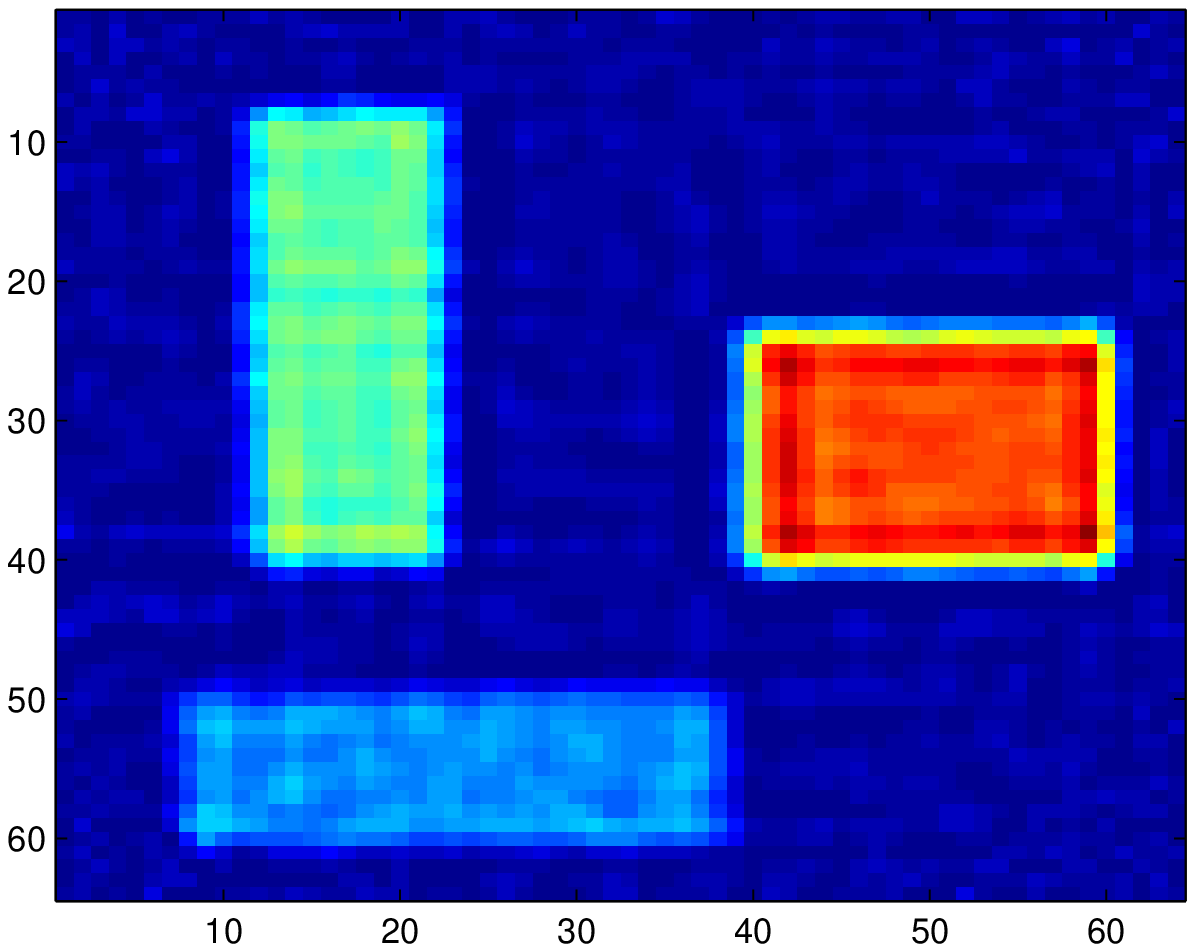}	\\
	\includegraphics[width=2in]{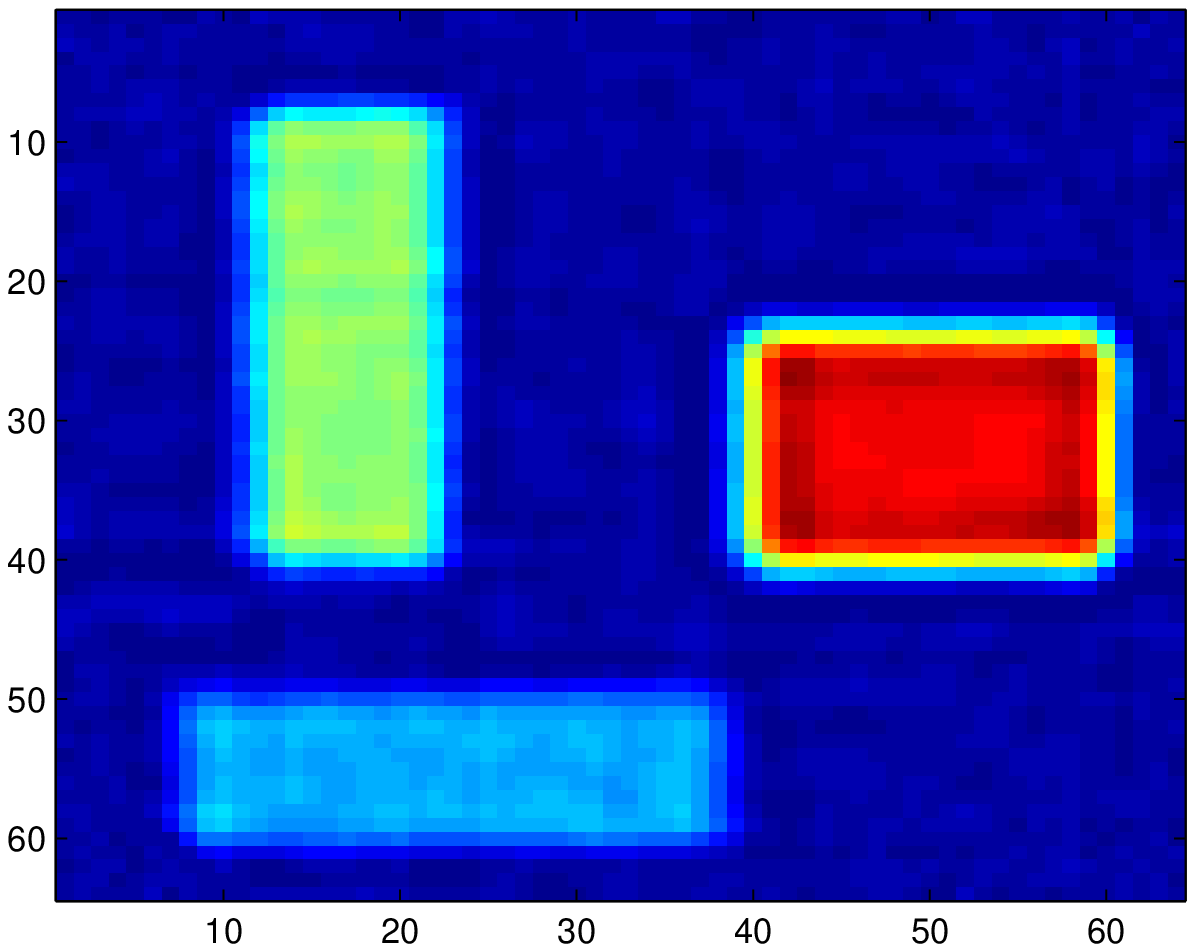} & \includegraphics[width=2in]{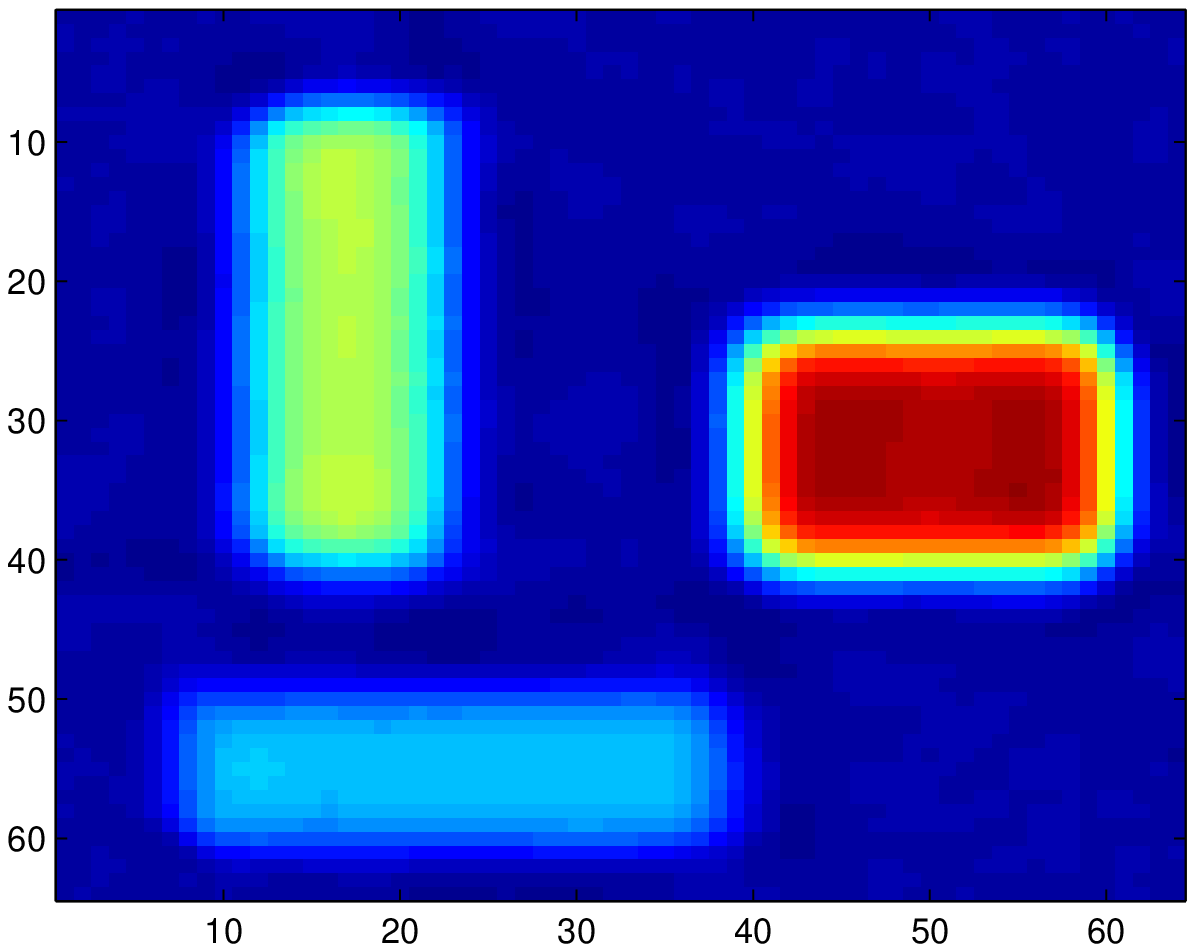}	\\
\end{array}
$$
\end{center}
\caption{The true PET image (top) and the recovered images with penalties $\mu= 0$, $10^{-7}$, $10^{-6}$, and $10^{-5}$.}
\label{fig:pet}
\end{figure}

Table~\ref{table:pet} compares the run times of the CPU and GPU implementations for a simulated PET image \citep{Roland07PET}. The image as depicted in the top of Figure~\ref{fig:pet} has $p=64 \times 64=4,096$ pixels and is interrogated by $d=2,016$ detectors. Overall we see a 43- to 53-fold reduction in run times with the GPU implementation. Figure~\ref{fig:pet} displays the true image and the estimated images under penalties of $\mu=0$, $10^{-5}$, $10^{-6}$, and $10^{-7}$. Without penalty ($\mu=0$), the algorithm fails to converge in 100,000 iterations.

\begin{algorithm}
\begin{algorithmic}
\STATE Scale $\bf{E}$ to have unit $l_1$ column norms.
\STATE Compute $|{\cal N}_j| = \sum_{k: \{j,k\} \in {\cal N}} 1$ and $a_j -2 \mu |{\cal N}_j|$ for all $1 \le j \le p$.
\STATE Initialize: $\lambda_{0j} \leftarrow 1$, $j=1,\ldots,p$.
\REPEAT
\STATE $z_{nij} \leftarrow (y_i e_{ij}\lambda_{nj})/(\sum_k e_{ik} \lambda_{nk})$ for all $1 \le i \le d$, $1 \le j \le p$
\FOR{$j=1$ to $p$}
\STATE $b_{nj} \leftarrow \mu(|{\cal N}_j|\lambda_{nj}+\sum_{k \in {\cal N}_j} \lambda_{nk})-1$
\STATE $c_{nj} \leftarrow \sum_i z_{nij}$
\STATE $\lambda_{n+1,j} \leftarrow (-b_{nj}-\sqrt{b_{nj}^{2}-4a_jc_{nj}})/(2a_j)$
\ENDFOR
\UNTIL{convergence occurs}
\end{algorithmic}
\caption{(PET Image Recovering) Given the coefficient matrix $\mathbf{E} \in \mathbb{R}_+^{d \times p}$, coincident counts ${\bf y} = (y_1,\ldots,y_d) \in \mathbf{Z}_+^d$, and roughness parameter $\mu>0$, find the intensity vector $\bflambda=(\lambda_1,\ldots,\lambda_p) \in \mathbb{R}_+^p$ that maximizes the objective function~(\ref{eqn:PET-objpen}).}
\label{algo:PET}
\end{algorithm}

\begin{sidewaystable}
\begin{center}
\begin{tabular}{crrrrrrcrrrc}
\toprule
& \multicolumn{3}{c}{CPU} & \multicolumn{4}{c}{GPU} & \multicolumn{4}{c}{QN(10) on CPU} \\
 \cmidrule(lr){2-4} \cmidrule(lr){5-8} \cmidrule(lr){9-12}
Penalty $\mu$ & Iters & Time  & Function & Iters & Time & Function & Speedup & Iters & Time & Function & Speedup	\\
\midrule
0 & 100000   & 14790    & -7337.152765  & 100000   & 282      & -7337.153387 & 52 & 6549 & 2094 & -7320.100952 & n/a \\
$10^{-7}$ & 24457    & 3682     & -8500.083033  & 24457    & 70       & -8508.112249 & 53 & 251 & 83 & -8500.077057 & 44 \\
$10^{-6}$ & 6294     & 919      & -15432.45496  & 6294     & 18       & -15432.45586 & 51 & 80 & 29 & -15432.45366 & 32 \\
$10^{-5}$ & 589      & 86       & -55767.32966  & 589      & 2        & -55767.32970 & 43 & 19 & 9 & -55767.32731 & 10 \\
\bottomrule
\end{tabular}
\end{center}
\caption{Comparison of run times (in seconds) for a PET imaging problem on the simulated data in \cite{Roland07PET}. The image has $p=64 \times 64=4,096$ pixels and is interrogated by $d=2,016$ detectors. The columns labeled Function refer to the converged value of the objective function. The results under the heading {\tt $QN(10)$ on CPU} invoke quasi-Newton acceleration \cite{ZhouAlexanderLange09QN} with 10 secant conditions.}
\label{table:pet}
\end{sidewaystable}

\subsection{Multidimensional Scaling}

Multidimensional scaling (MDS) was the first statistical application of the MM principle \cite{deLeeuw76MDS,deLeeuwHeiser77MDS}. MDS represents $q$ objects as faithfully as possible in $p$-dimensional space given a nonnegative weight $w_{ij}$ and a nonnegative dissimilarity measure $y_{ij}$ for each pair of objects $i$ and $j$. If $\bftheta^i \in \mathbb{R}^p$ is the position of object $i$, then the $p \times q$ parameter matrix $\bftheta = (\bftheta^1, \ldots, \bftheta^q)$ is estimated by minimizing the stress function
\begin{eqnarray}
f(\bftheta)	&=& \sum_{1 \le i < j \le q} w_{ij} (y_{ij} - \|\bftheta^i-\bftheta^j\|)^2 \label{eqn:stress} \\
&=& \sum_{i < j} w_{ij} y_{ij}^2 - 2 \sum_{i < j} w_{ij} y_{ij} \|\bftheta^i - \bftheta^j\| 
 + \sum_{i < j} w_{ij} \|\bftheta^i - \bftheta^j\|^2, \nonumber
\end{eqnarray}
where $\|\bftheta^i - \bftheta^j\|$ is the Euclidean distance between $\bftheta^i$ and $\bftheta^j$. The stress function~(\ref{eqn:stress}) is invariant under translations, rotations, and reflections of $\mathbb{R}^p$. To avoid translational and rotational ambiguities, we take $\bftheta^1$ to be the origin and the first $p-1$ coordinates of $\bftheta^2$ to be 0. Switching the sign of 
$\theta^2_{p}$ leaves the stress function invariant. Hence, convergence to one member of a pair of reflected minima immediately determines the other member. 

Given these preliminaries, we now review the derivation of the MM algorithm presented in \citep{Lange00OptTrans}. Because we want to minimize the stress, we majorize it. The middle term in the stress~(\ref{eqn:stress}) is majorized by the Cauchy-Schwartz inequality
\begin{eqnarray*}
- \|\bftheta^i - \bftheta^j\|  & \le & - \frac{(\bftheta^i - \bftheta^j)^t (\bftheta^i_n - \bftheta^j_n)}
{\|\bftheta^i_n - \bftheta^j_n\|}.
\end{eqnarray*}
To separate the parameters in the summands of the third term of the stress, we invoke the convexity of the Euclidean norm $\|\cdot\|$ and the square function $s^2$. These maneuvers yield
\begin{eqnarray*}
\|\bftheta^i-\bftheta^j\|^2 & = & \Big\|\frac{1}{2}\Big[2\bftheta^i - (\bftheta^i_n+\bftheta^j_n)\Big]
- \frac{1}{2}\Big[2\bftheta^j -(\bftheta^j_n+\bftheta^j_n)\Big]\Big\|^2 \\
& \le &  2 \Big\|\bftheta^i - \frac 12 (\bftheta^i_n + \bftheta^j_n)\Big\|^2 
+ 2 \Big\|\bftheta^j - \frac 12 (\bftheta^i_n+\bftheta^j_n)\Big\|^2.
\end{eqnarray*}
Assuming that $w_{ij}=w_{ji}$ and $y_{ij}=y_{ji}$, the surrogate function therefore becomes
\begin{eqnarray*}
g(\bftheta \mid \bftheta_n) &=& 2 \sum_{i < j} w_{ij} \left[ \Big\|\bftheta^i - \frac 12 (\bftheta^i_n + \bftheta^j_n)\Big\|^2 - \frac{y_{ij} (\bftheta^i)^t (\bftheta^i_n - \bftheta^j_n)}{\|\bftheta^i_n-\bftheta^j_n\|}\right] \\
&   &  + 2 \sum_{i<j}w_{ij} \left[ \Big\|\bftheta^j - \frac 12 (\bftheta^i_n+\bftheta^j_n)\Big\|^2 + \frac{y_{ij} (\bftheta^j)^t(\bftheta^i_n-\bftheta^j_n)}{\|\bftheta^i_n-\bftheta^j_n\|} \right]	\\
& = & 2 \sum_{i=1}^q \sum_{j \ne i} \left[ w_{ij} \Big\|\bftheta^i - \frac 12 (\bftheta^i_n+\bftheta^j_n)\Big\|^2 - \frac{w_{ij} y_{ij} (\bftheta^i)^t(\bftheta^i_n-\bftheta^j_n)}{\|\bftheta^i_n-\bftheta^j_n\|} \right]
\end{eqnarray*}
up to an irrelevant constant. Setting the gradient of the surrogate equal to the $\mathbf{0}$ vector produces the parallel updates
\begin{eqnarray*}
\theta^{i}_{n+1,k} & = & \frac{\sum_{j \ne i} \left[ \frac{w_{ij}y_{ij}(\theta^{i}_{nk}-\theta^{j}_{nk})}
{\|\bftheta^i_n-\bftheta^j_n\|} + w_{ij} (\theta^{i}_{nk} + \theta^{j}_{nk}) \right]}{2 \sum_{j \ne i} w_{ij}}
\end{eqnarray*}
for all movable parameters $\theta^{i}_{k}$. 

Algorithm \ref{algo:mds} summarizes the parallel organization of the steps. Again the matrix multiplications $\mathbf{\Theta}_{n}^t \mathbf{\Theta}_n$ and $\mathbf{\Theta}_n(\mathbf{W}-\mathbf{Z}_n)$ can be taken care of by the CUBLAS library \cite{cublas-book08}. The remaining steps of the algorithm are conducted by easily written parallel code.

\begin{figure}
\begin{center}
\includegraphics[width=4.5in]{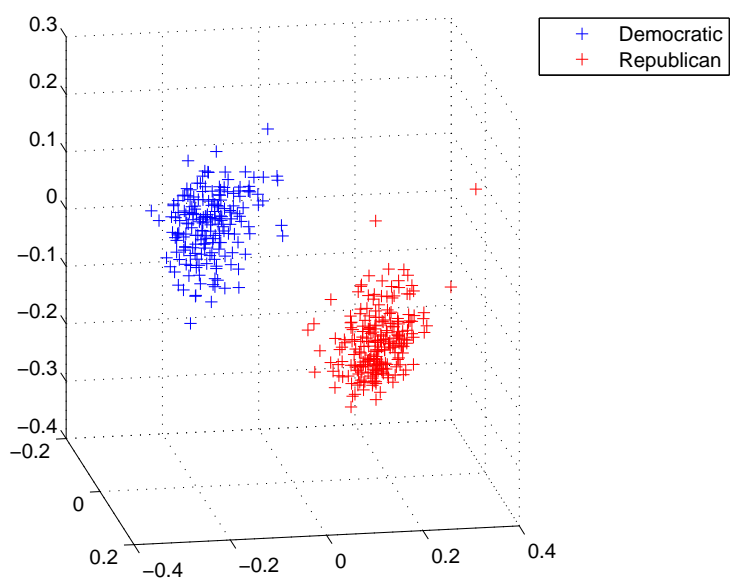}
\end{center}
\caption{Display of the MDS results with $p=3$ coordinates on the 2005 House of Representatives roll call data.}
\label{fig:mds}
\end{figure}

Table \ref{table:mds} compares the run times in seconds for MDS on the 2005 United States House of Representatives roll call votes. The original data consist of the 671 roll calls made by 401 representatives. We refer readers to the reference \citep{Diaconis08Horseshoes} for a careful description of the data and how the MDS input $401 \times 401$ distance matrix is derived. The weights $w_{ij}$ are taken to be 1. In our notation, the number of objects (House Representatives) is $q=401$. Even for this relatively small dataset, we see a 27--48 fold reduction in total run times, depending on the projection dimension $p$. Figure~\ref{fig:mds} displays the results in $p=3$ dimensional space. The Democratic and Republican members are clearly separated. For $p=30$, the algorithm fails to converge within 100,000 iterations.

Although the projection of points into $p>3$ dimensional spaces may sound artificial, there are situations where this is standard practice. First, MDS is foremost a dimension reduction tool, and it is desirable to keep $p>3$ to maximize explanatory power. Second, the stress function tends to have multiple local minima in low dimensions \cite{Groenen96Tunneling}. A standard optimization algorithm like MM is only guaranteed to converge to a local minima of the stress function. As the number of dimensions increases, most of the inferior modes disappear. One can formally demonstrate that the stress has a unique minimum when $p=q-1$ \cite{deLeeuw93,Groenen96Tunneling}. In practice, uniqueness can set in well before $p$ reaches $q-1$. In the recent work \cite{ZhouLange09Annealing}, we propose a ``dimension crunching" technique that increases the chance of the MM algorithm converging to the global minimum of the stress function. In dimension crunching, we start optimizing the stress in a Euclidean space $\mathbb{R}^m$ with $m>p$. The last $m-p$ components of each column $\bftheta^i$ are gradually subjected to stiffer and stiffer penalties. In the limit as the penalty tuning parameter tends to $\infty$, we recover the global minimum of the stress in $\mathbb{R}^p$. This strategy inevitably incurs a computational burden when $m$ is large, but the MM+GPU combination comes to the rescue.

\begin{algorithm}
\begin{algorithmic}
\STATE Precompute: $x_{ij} \leftarrow w_{ij} y_{ij}$ for all $1 \le i,j \le q$
\STATE Precompute: $w_{i\cdot} \leftarrow \sum_j w_{ij}$ for all $1 \le i \le q$
\STATE Initialize: Draw $\theta^{i}_{0k}$ uniformly on [-1,1] for all $1 \le i \le q$, $1 \le k \le p$
\REPEAT
\STATE Compute $\mathbf{\Theta}_n^t \mathbf{\Theta}_n$
\STATE $d_{nij} \leftarrow \{\mathbf{\Theta}_n^t \mathbf{\Theta}_n\}_{ii} + \{\mathbf{\Theta}_n^t \mathbf{\Theta}_n\}_{jj} - 2 \{\mathbf{\Theta}_n^t \mathbf{\Theta}_n\}_{ij}$ for all $1 \le i, j \le q$
\STATE $z_{nij} \leftarrow x_{ij} / d_{nij}$ for all $1 \le i \ne j \le q$
\STATE $z_{ni\cdot} \leftarrow \sum_j z_{nij}$ for all $1 \le i \le q$
\STATE Compute $\mathbf{\Theta}_n (\mathbf{W}-\mathbf{Z}_n)$
\STATE $\theta_{n+1,k}^i \leftarrow [\theta_{nk}^i (w_{i\cdot}+z_{ni\cdot}) + \{\mathbf{\Theta}_n (\mathbf{W}-\mathbf{Z}_n)\}_{ik}] / (2w_{i\cdot})$ for all $1 \le i \le p$, $1 \le k \le q$
\UNTIL{convergence occurs}
\end{algorithmic}
\caption{(MDS) Given weights $\mathbf{W}$ and distances $\mathbf{Y} \in \mathbb{R}^{q \times q}$, find the matrix $\mathbf{\Theta} = [\bftheta^1, \ldots, \bftheta^q] \in \mathbb{R}^{p \times q}$ which minimizes the stress~(\ref{eqn:stress}).}
\label{algo:mds}
\end{algorithm}

\begin{sidewaystable}
\begin{center}
\begin{tabular}{crrrrrrcrrrc}
\toprule
& \multicolumn{3}{c}{CPU} & \multicolumn{4}{c}{GPU} & \multicolumn{4}{c}{QN(20) on CPU} \\
 \cmidrule(lr){2-4} \cmidrule(lr){5-8} \cmidrule(lr){9-12}
Dim-$p$ & Iters & Time  & Stress & Iters & Time & Stress & Speedup & Iters & Time & Stress & Speedup	\\
\midrule
2 & 3452     & 43       & 198.5109307  & 3452     & 1        & 198.5109309 & 43 & 530 & 16 & 198.5815072 & 3 \\
3 & 15912    & 189      & 95.55987770  & 15912    & 6        & 95.55987813 & 32 & 1124 & 38 & 92.82984196 & 5 \\
4 & 15965    & 189      & 56.83482075  & 15965    & 7        & 56.83482083 & 27 & 596 & 18 & 56.83478026 & 11 \\
5 & 24604    & 328      & 39.41268434  & 24604    & 10       & 39.41268444 & 33 & 546 & 17 & 39.41493536 & 19 \\
10 & 29643    & 441      & 14.16083986  & 29643    & 13       & 14.16083992 & 34 & 848 & 35 & 14.16077368 & 13 \\
20 & 67130    & 1288     & 6.464623901  & 67130    & 32       & 6.464624064 & 40 & 810 & 43 & 6.464526731 & 30  \\
30 & 100000   & 2456     & 4.839570118  & 100000   & 51       & 4.839570322 & 48 & 844 & 54 & 4.839140671 & n/a \\
\bottomrule
\end{tabular}
\end{center}
\caption{Comparison of run times (in seconds) for MDS on the 2005 House of Representatives roll call data. The number of points (representatives) is $q=401$. The results under the heading {\tt $QN(20)$ on CPU} invoke the quasi-Newton acceleration \cite{ZhouAlexanderLange09QN} with 20 secant conditions.}
\label{table:mds}
\end{sidewaystable}

\section{Discussion}
\label{discussion_section}

The rapid and sustained increases in computing power over the last half century have transformed statistics. Every advance has encouraged statisticians to attack harder and more sophisticated problems. We tend to take the steady march of computational efficiency for granted, but there are limits to a chip's clock speed, power consumption, and logical complexity. Parallel processing via GPUs is the technological
innovation that will power ambitious statistical computing in the coming decade. Once the limits of parallel processing are reached, we may see quantum computers take off. In the meantime statisticians should learn how to harness GPUs productively.

We have argued by example that high-dimensional optimization is driven by parameter and data separation. It takes both to exploit the parallel capabilities of GPUs.  Block relaxation and the MM algorithm often generate ideal parallel algorithms. In our opinion the MM algorithm is the more versatile of the two generic strategies. Unfortunately, block relaxation does not accommodate constraints well and may generate sequential rather than parallel updates. Even when its updates are parallel, they may not be data separated. The EM algorithm is one of the most versatile tools in the statistician's toolbox. The MM principle generalizes the EM algorithm and shares its positive features. Scoring and Newton's methods become impractical in high dimensions. Despite these arguments in favor of MM algorithms, one should always keep in mind hybrid algorithms such as the one we implemented for NNMF.

Although none of our data sets is really large by today's standards, they do demonstrate that a good GPU implementation can easily achieve one to two orders of magnitude improvement over a single CPU core.  Admittedly, modern CPUs come with 2 to 8 cores, and distributed computing over CPU-based clusters remains an option. But this alternative also carries a hefty price tag. The NVIDIA GTX280 GPU on which our examples were run drives 240 cores at a cost of several hundred dollars. High-end computers with 8 or more CPU nodes cost thousands of dollars. It would take 30 CPUs with 8 cores each to equal a single GPU at the same clock rate. Hence, GPU cards strike an effective and cost efficient balance.

The simplicity of MM algorithms often comes at a price of slow (at best linear) convergence. Our MDS, NNMF, and PET (without penalty) examples are cases in point. Slow convergence is a concern as statisticians head into an era dominated by large data sets and high-dimensional models. Think about the scale of the Netflix data matrix. The speed of any iterative algorithm is determined by both the computational cost per iteration and the number of iterations until convergence. GPU implementation reduces the first cost. Computational statisticians also have a bag of software tricks to decrease the number of iterations \cite{MengRubin93ECM,Jamshidian93ConjGrad,LiuRubin94ECME,Lange95QuasiNewton,Jamshidian97,MengVanDyk97,Varadhan08SQUAREM}. For instance, the recent paper~\cite{ZhouAlexanderLange09QN} proposes a quasi-Newton acceleration scheme particularly suitable for high-dimensional problems. The scheme is off-the-shelf and broadly applies to any search algorithm defined by a smooth algorithm map. The acceleration requires only modest increments in storage and computation per iteration. Tables~\ref{table:pet} and~\ref{table:mds} also list the results of this quasi-Newton acceleration of the CPU implementation for the MDS and PET examples. As the tables make evident, quasi-Newton acceleration significantly reduces the number of iterations until convergence. The accelerated algorithm always locates a better mode while cutting run times compared to the unaccelerated algorithm. We have tried the quasi-Newton acceleration on our GPU hardware with mixed results. We suspect that the lack of full double precision on the GPU is the culprit. When full double precision becomes widely available, the combination of GPU hardware acceleration and algorithmic software acceleration will be extremely potent.

Successful acceleration methods will also facilitate attacking another nagging problem in computational statistics, namely multimodality. No one knows how often statistical inference is fatally flawed because a standard optimization algorithm converges to an inferior mode. The current remedy of choice is to start a search algorithm from multiple random points. Algorithm acceleration is welcome because the number of starting points can be enlarged without an increase in computing time. As an alternative to multiple starting points, our recent paper~\cite{ZhouLange09Annealing} suggests modifications of several standard MM algorithms that increase the chance of locating better modes. These simple modifications all involve variations on deterministic annealing~\cite{Ueda98DAEM}.

Our treatment of simple classical examples should not hide the wide applicability of the powerful MM+GPU combination. A few other candidate applications include penalized estimation of haplotype frequencies in genetics \cite{ayers08}, construction of biological and social networks under a random multigraph model \cite{ranola10}, and data mining with a variety of models related to the multinomial distribution \cite{zhou10}. Many mixture models will benefit as well from parallelization, particularly in assigning group memberships.  Finally, parallelization is hardly limited to optimization. We can expect to see many more GPU applications in MCMC sampling. Given the computationally intensive nature of MCMC, the ultimate payoff may even be higher in the Bayesian setting than in the frequentist setting.  Of course realistically, these future triumphs will require a great deal of thought, effort, and education.  There is usually a desert to wander and a river to cross before one reaches the promised land.

\section*{Acknowledgements}
M.S. acknowledges support from NIH grant R01 GM086887. K.L. was supported by United States Public Health Service grants GM53275 and MH59490.


\begin{thebibliography}{}

%
%
%
%

\bibitem[\protect\citeauthoryear{Ayers, K.~L. and Lange, K.~L.}{2008}]{ayers08}
\textsc{Ayers, K.~L.} and \textsc{Lange, K.~L} (2008). 
Penalized estimation of haplotype frequencies. 
\textit{Bioinformatics} \textbf{24} 1596--1602.

\bibitem[\protect\citeauthoryear{Berry, et al}{2007}]{Berry07NNMF}
\textsc{Berry, M.~W.}, \textsc{Browne, M.}, \textsc{Langville, A.~N.}, \textsc{Pauca, V.~P.}, and \textsc{Plemmons, R.~J.} (2007).
Algorithms and applications for approximate nonnegative matrix
  factorization.
\textit{Comput. Statist. Data Anal.} \textbf{52} 155--173.
\MR{2409971}

\bibitem[\protect\citeauthoryear{Buckner, et al}{2009}]{buckner09}
\textsc{Buckner, J.}, \textsc{Wilson J.}, \textsc{Seligman, M.}, \textsc{Athey, B.}, \textsc{Watson, S.} and \textsc{Meng, F.} (2009)
The gputools package enables GPU computing in R. 
\textit{Bioinformatics} \textbf{22} btp608.

\bibitem[\protect\citeauthoryear{de Leeuw, J.}{unpublished manuscript}]{deLeeuw93}
\textsc{de~Leeuw, J.}
Fitting distances by least squares.
\textit{unpublished manuscript}.

\bibitem[\protect\citeauthoryear{de Leeuw, J. and Heiser, W.~J.}{1977}]{deLeeuwHeiser77MDS}
\textsc{de~Leeuw, J.} and \textsc{Heiser, W.~J.} (1977).
Convergence of correction matrix algorithms for multidimensional scaling.
\textit{Geometric Representations of Relational Data}, 133--145. Mathesis Press, Ann Arbor, MI.

\bibitem[\protect\citeauthoryear{de Leeuw, J.}{1977}]{deLeeuw76MDS}
\textsc{de~Leeuw, J.} (1977).
Applications of convex analysis to multidimensional scaling.
\textit{Recent developments in statistics ({P}roc. {E}uropean
  {M}eeting {S}tatisticians, {G}renoble, 1976)}, 133--145. North-Holland,
  Amsterdam.

\bibitem[\protect\citeauthoryear{Dempster, et al}{1977}]{Dempster77EM}
\textsc{Dempster, A.~P.}, \textsc{Laird, N.~M.}, and \textsc{Rubin, D.~B.} (1977).
Maximum likelihood from incomplete data via the {EM} algorithm. (with discussion)
\textit{J. Roy. Statist. Soc. Ser. B} \textbf{39} 1--38.
\MR{0501537}

\bibitem[\protect\citeauthoryear{Diaconis, et al}{2008}]{Diaconis08Horseshoes}
\textsc{Diaconis, P.}, \textsc{Goel, S.}, and \textsc{Holmes, S.} (2008).
Horseshoes in multidimensional scaling and local kernel methods.
\textit{Annals of Applied Statistics} \textbf{2} 777--807.

\bibitem[\protect\citeauthoryear{Groenen, P.~J.~F. and Heiser, W.~J.}{1996}]{Groenen96Tunneling}
\textsc{Groenen, P.~J.~F.} and \textsc{Heiser, W.~J.} (1996).
The tunneling method for global optimization in multidimensional
  scaling.
\textit{Pshychometrika} \textbf{61} 529--550.


\bibitem[\protect\citeauthoryear{Jamshidian, M. and Jennrich, R.~I.}{1993}]{Jamshidian93ConjGrad}
\textsc{Jamshidian, M.} and \textsc{Jennrich, R.~I.} (1993).
Conjugate gradient acceleration of the {EM} algorithm.
\textit{J. Amer. Statist. Assoc.} \textbf{88} 221--228.
\MR{1212487}

\bibitem[\protect\citeauthoryear{Jamshidian, M. and Jennrich, R.~I.}{1997}]{Jamshidian97}
\textsc{Jamshidian, M} and \textsc{Jennrich, R.~I.} (1997).
Acceleration of the {EM} algorithm by using quasi-{N}ewton methods.
\textit{J. Roy. Statist. Soc. Ser. B} \textbf{59} 569--587.
\MR{1452026}

\bibitem[\protect\citeauthoryear{Koren, Y. et al}{2009}]{KorBel09}
\textsc{Koren, Y}, \textsc{Bell, R.}, and \textsc{Volinsky, C.} (2009).
Matrix factorization techniques for recommender systems.
\textit{Computer} \textbf{42} 30--37.


\bibitem[\protect\citeauthoryear{Lange, K.~L. and Carson, R.}{1984}]{Lange84PET}
\textsc{Lange, K.~L.} and \textsc{Carson, R.} (1984).
{EM} reconstruction algorithms for emission and transmission
  tomography.
\textit{J. Comput. Assist. Tomogr.} \textbf{8} 306--316.

\bibitem[\protect\citeauthoryear{Lange, K.~L.}{1995}]{Lange95QuasiNewton}
\textsc{Lange, K.~L.} (1995).
A quasi-{N}ewton acceleration of the {EM} algorithm.
\textit{Statist. Sinica} \textbf{5} 1--18.
\MR{1329286}

\bibitem[\protect\citeauthoryear{Lange, K.~L.}{2004}]{Lange04Optm}
\textsc{Lange, K.~L.} (2004).
\textit{Optimization}.
Springer-Verlag, New York.
\MR{2072899}

\bibitem[\protect\citeauthoryear{Lange, et al}{2000}]{Lange00OptTrans}
\textsc{Lange, K.~L.}, \textsc{Hunter, D.~R.}, and \textsc{Yang, I.} (2000).
Optimization transfer using surrogate objective functions. (with discussion)
\textit{J. Comput. Graph. Statist.} \textbf{9} 1--59.
\MR{1819865}

\bibitem[\protect\citeauthoryear{Lee et al}{2009}]{holmes-tech-report}
\textsc{Lee, A.}, \textsc{Yan, C.}, \textsc{Giles, M.~B.}, \textsc{Doucet, A.}, and \textsc{Holmes, C.~C.} (2009).
On the utility of graphics cards to perform massively parallel
simulation of advanced Monte Carlo methods.
\textit{Technical report}, Department of Statistics, Oxford University.

\bibitem[\protect\citeauthoryear{Lee and Seung}{1999}]{LeeSeung99NNMF}
\textsc{Lee, D.~D.} and \textsc{Seung, H.~S.} (1999).
Learning the parts of objects by non-negative matrix factorization.
\textit{Nature} \textbf{401} 788--791.

\bibitem[\protect\citeauthoryear{Lee and Seung}{2001}]{LeeSeung01NNMFAlgo}
\textsc{Lee, D.~D.} and \textsc{Seung, H.~S.} (2001).
Algorithms for non-negative matrix factorization.
\textit{NIPS}, pages 556--562, MIT Press.

\bibitem[\protect\citeauthoryear{Liu, C. and Rubin, D.~B.}{1994}]{LiuRubin94ECME}
\textsc{Liu, C.} and \textsc{Rubin, D.~B.} (1994).
The {ECME} algorithm: a simple extension of {EM} and {ECM} with
  faster monotone convergence.
\textit{Biometrika} \textbf{81} 633--648.
\MR{1326414}

\bibitem[\protect\citeauthoryear{McLachlan, G.~J. and Krishnan, T.}{2008}]{McLachlan08EMBook}
\textsc{McLachlan, G.~J.} and \textsc{Krishnan, T.} (2008).
\textit{The {EM} algorithm and extensions}.
Wiley-Interscience [John Wiley \& Sons], Hoboken, NJ, second edition.
\MR{2392878}

\bibitem[\protect\citeauthoryear{Meng, X.~L. and Rubin, D.~B.}{1993}]{MengRubin93ECM}
\textsc{Meng, X.~L.} and \textsc{Rubin, D.~B.} (1993).
Maximum likelihood estimation via the {ECM} algorithm: a general framework.
\textit{Biometrika} \textbf{80} 267--278.
\MR{1243503}

\bibitem[\protect\citeauthoryear{Meng, X.~L. and van Dyk, D.}{1993}]{MengVanDyk97}
\textsc{Meng, X.~L.} and \textsc{van Dyk, D.} (1997).
The {EM} algorithm---an old folk-song sung to a fast new tune. (with discussion)
\textit{J. Roy. Statist. Soc. Ser. B}, 59(3):511--567.
\MR{1452025}

\bibitem{MITCBCL}
{MIT} center for biological and computational learning.
\newblock CBCL Face~Database \#1,
\url{http://www.ai.mit.edu/projects/cbcd}.

\bibitem{cublas-book08}
\textsc{NVIDIA} (2008).
NVIDIA CUBLAS Library. 

\bibitem{cuda-book}
\textsc{NVIDIA} (2008).
{NVIDIA CUDA Compute Unified Device Architecture: Programming Guide Version 2.0}.

\bibitem[\protect\citeauthoryear{Owens, et al}{2007}]{Owens:2007:ASO}
\textsc{Owens, J.~D.}, \textsc{Luebke, D.}, \textsc{Govindaraju, N.}, \textsc{Harris, M.}, \textsc{Kr{\"{u}}ger, J.}, 
  \textsc{Lefohn, A.~E.}, and \textsc{Purcell, T.~J.} (2007).
A survey of general-purpose computation on graphics hardware.
\textit{Computer Graphics Forum} \textbf{26} 80--113.

\bibitem[\protect\citeauthoryear{Ranola, et al}{2010}]{ranola10}
\textsc{Ranola, J.M.}, \textsc{Ahn, S.}, \textsc{Sehl, M.E.}, \textsc{Smith, D.J.} and \textsc{Lange, K.~L.} (2010) 
A Poisson model for random multigraphs.
\textit{unpublished manuscript}.

\bibitem[\protect\citeauthoryear{Roland, et al}{2007}]{Roland07PET}
\textsc{Roland, C.}, \textsc{Varadhan, R.}, and \textsc{Frangakis, C.~E.} (2007).
Squared polynomial extrapolation methods with cycling: an application
  to the positron emission tomography problem.
\textit{Numer. Algorithms} \textbf{44} 159--172.
\MR{2334694}


\bibitem[\protect\citeauthoryear{Silberstein et al}{2008}]{silberstein08}
\textsc{Silberstein, M.}, \textsc{Schuster, A.}, \textsc{Geiger, D.}, \textsc{Patney, A.}, and \textsc{Owens, J.~D.} (2008).
Efficient computation of sum-products on {GPU}s through
  software-managed cache.
\textit{Proceedings of the 22nd Annual International
  Conference on Supercomputing}, pages 309--318, ACM.

\bibitem[\protect\citeauthoryear{Sinnott-{A}rmstrong, et al}{2009}]{sinnott-armstrong09}
\textsc{Sinnott-{A}rmstrong, N.~A.}, \textsc{Greene, C.~S.}, \textsc{Cancare, F.}, and \textsc{Moore, J.~H.} (2009).
Accelerating epistasis analysis in human genetics with consumer
  graphics hardware.
\textit{{BMC} Research Notes} \textbf{2} 149.

\bibitem[\protect\citeauthoryear{Suchard and Rambaut}{2009}]{suchard09}
\textsc{Suchard, M.~A.} and \textsc{Rambaut, A.} (2009).
Many-core algorithms for statistical phylogenetics.
\textit{Bioinformatics} \textbf{25} 1370--1376.

\bibitem[\protect\citeauthoryear{Tibbits et al}{2009}]{tibbets09}
\textsc{Tibbits, M.~M.}, \textsc{Haran, M.}, and \textsc{Liechty, J.~C.} (2009).
Parallel multivariate slice sampling.
\textit{Statistics and Computing}, to appear.

\bibitem[\protect\citeauthoryear{Ueda, N. and Nakano, R.}{1998}]{Ueda98DAEM}
\textsc{Ueda, N.} and \textsc{Nakano, R.} (1998).
Deterministic annealing {EM} algorithm.
\textit{Neural Networks} \textbf{11} 271 -- 282.

\bibitem[\protect\citeauthoryear{Varadhan, R. and Roland, C.}{2008}]{Varadhan08SQUAREM}
\textsc{Varadhan, R.} and \textsc{Roland, C.} (2008).
\newblock Simple and globally convergent methods for accelerating the
  convergence of any {EM} algorithm.
\textit{Scand. J. Statist.} \textbf{35} 335--353.
\MR{2418745}

\bibitem[\protect\citeauthoryear{Vardi, et al}{1985}]{Vardi85PET}
\textsc{Vardi, Y.}, \textsc{Shepp, L.~A.}, and \textsc{Kaufman, L.} (1985).
\newblock A statistical model for positron emission tomography. (with discussion)
\textit{J. Amer. Statist. Assoc.} \textbf{80} 8--37.
\MR{786595}

\bibitem[\protect\citeauthoryear{Wu and Lange}{2009}]{WuLange09EMMM}
\textsc{Wu, T.~T.} and \textsc{Lange, K.~L.} (2009).
\newblock The {MM} alternative to {EM}.
\textit{Stat. Sci.}, in press.

\bibitem[\protect\citeauthoryear{Zhou, H. and Lange, K.~L.}{2009}]{ZhouLange09Annealing}
\textsc{Zhou, H.} and \textsc{Lange K.~L.} (2009).
\newblock On the bumpy road to the dominant mode.
\textit{Scandinavian Journal of Statistics}, in press.

\bibitem[\protect\citeauthoryear{Zhou, et al}{2009}]{ZhouAlexanderLange09QN}
\textsc{Zhou, H.}, \textsc{Alexander, D.}, and \textsc{Lange, K.~L.} (2009).
\newblock A quasi-newton acceleration for high-dimensional optimization
  algorithms.
\textit{Statistics and Computing}, DOI:10.1007/s11222-009-9166-3.

\bibitem[\protect\citeauthoryear{Zhou, H. and Lange, K.~L.}{2009}]{zhou10}
\textsc{Zhou, H.} and \textsc{Lange, K.~L.} (2009).
\newblock MM algorithms for some discrete multivariate distributions.
\textit{J Computational Graphical Stat}, in press.

\end{thebibliography}
\end{document}